\documentclass[11pt,a4paper]{article}
\usepackage{amsfonts}
\usepackage{amsbsy}
\usepackage{amssymb}
\usepackage{amsmath}
\usepackage{mathrsfs}
\usepackage{latexsym}
\usepackage{color}
\usepackage{revsymb}

\def\ihalf{\textstyle{\frac{i}{2}}}

\def\onehalf{\textstyle{\frac{1}{2}}}

\def\D{{\mathcal D}{}}

\def\Abol{{\stackrel{~\circ}{A}}{}}

\def\Rbol{{\stackrel{\circ}{R}}{}}
\def\Lbol{{\stackrel{\circ}{\mathcal L}}{}}

\def\Sbol{{\stackrel{\circ}{S}}{}}
\def\jbol{{\stackrel{\circ}{\jmath}}{}}

\def\Aw{{\stackrel{~\bullet}{A}}{}}

\def\Rw{{\stackrel{\bullet}{R}}{}}

\def\jw{{\stackrel{\bullet}{\jmath}}{}}
\def\iw{{\stackrel{\bullet}{\imath}}{}}
\def\tw{{\stackrel{\bullet}{t}}{}}

\def\Lw{{\stackrel{\bullet}{\mathcal L}}{}}
\def\Tw{{\stackrel{\bullet}{T}}{}}

\def\Kw{{\stackrel{\bullet}{K}}{}}

\def\Dw{{\stackrel{\bullet}{\mathcal D}}{}}

\def\sw{{\stackrel{\bullet}{S}}{}}

\def\siw{{\stackrel{\bullet}{\sigma}}{}}

\def\be{\begin{equation}}
\def\ee{\end{equation}}
\def\ba{\begin{eqnarray}}
\def\ea{\end{eqnarray}}

\begin{document}

\renewcommand{\thefootnote}{\fnsymbol{footnote}}
\noindent
{\Large \bf Inertia and gravitation in teleparallel gravity}
\vskip 0.7cm
\noindent
{\bf R. Aldrovandi, Tiago Gribl Lucas and J. G. Pereira}
\vskip 0.1cm \noindent
{\it Instituto de F\'{\i}sica Te\'orica},
{\it UNESP-Universidade Estadual Paulista} \\
{\it Caixa Postal 70532-2, 
01156-970 S\~ao Paulo, Brazil}

\vskip 0.8cm
\begin{quote}
{\bf Abstract.}~{\footnotesize Using the fact that teleparallel gravity allows a separation between gravitation and inertia, explicit expressions for the gravitational and the inertial energy-momentum densities are obtained. It is shown that, like all other fields of nature, gravitation alone has a tensorial energy-momentum density which in a general frame is conserved in the covariant sense. Together with the inertial energy-momentum density, they form a pseudotensor which is conserved in the ordinary sense. An analysis of the role played by the gravitational and the inertial densities in the computation of the total energy and momentum of gravity is presented.}
\end{quote}

\section{Introduction}
\setcounter{footnote}{0}
\renewcommand{\thefootnote}{\arabic{footnote}}

The definition of an energy-momentum density for the gravitational field is one of the oldest
and most controversial problems of gravitation. If gravity is a true field, it should have a well defined energy-momentum density \cite{Synge,bondi}. However, it is usually asserted that such a density cannot be defined because of the equivalence principle \cite{mtw}. In fact, in the context of general relativity all attempts to identify an energy-momentum density for the gravitational field leads to complexes that are not true tensors. The first of such attempts was made by Einstein, who proposed an expression which was simply the canonical expression obtained from Noether's theorem \cite{trautman}. Several other attempts have been made, leading to different expressions for this pseudotensor \cite{others}.

On the other hand, due to the fact that teleparallel gravity allows a separation between gravitation and inertia \cite{Einstein05}, it turns out possible in this theory to write down a tensorial expression for the gravitational energy-momentum density. The purpose of this paper is to explore further this property. We will proceed as follows. In section 2, for the sake of completeness, we present the fundamentals of Lorentz connections and frames. In section 3 we introduce the potential form of the gravitational field equations and discuss the usual conservation law of the gravitational energy-momentum density. In section 4 we consider the specific case of general relativity, and show why in this theory it is not possible to define a tensorial expression for the energy-momentum density of gravity. Then, in section 5 we discuss the reasons why teleparallel gravity is able to separate gravitation from inertia. Using this property, we introduce in section 6 the notion of inertial energy-momentum density, and show that the energy-momentum density of gravity alone---without inertia---is a true tensor which, like any other field in the presence of gravitation, in a general frame is conserved in the covariant sense. Finally, in section 7, it is shown that, together with the inertial energy-momentum density, they make up a pseudotensor which is conserved in the ordinary sense, and always yields the physically relevant result for the total energy and momentum of gravitation.

\section{Lorentz connections and frames}

We are going to use the Greek alphabet $(\mu, \nu, \rho, \dots = 0,1,2,3)$ to denote indices related to spacetime, and the first half of the Latin alphabet $(a,b,c, \dots = 0,1,2,3)$ to denote indices related to the tangent spaces, each one a Minkowski spacetime with metric $\eta_{ab} = \mathrm{diag}(+1,-1,-1,-1)$.
A basic ingredient of any gravitational theory is the spin connection $A_\mu$, a connection assuming values in the Lie algebra of the Lorentz group,
\be
A_\mu = \onehalf \, A^{ab}{}_\mu \, S_{ab},
\ee
with $S_{ab}$ a given representation of the Lorentz generators. The corresponding covariant derivative is given by the Fock-Ivanenko operator \cite{fi,dirac}
\be
\D_\mu = \partial_\mu -  \ihalf \, A^{ab}{}_\mu \, S_{ab}.
\ee
Acting on a Lorentz vector field $\phi^a$, for example, $S_{ab}$ is a matrix with entries
\[
(S_{ab})^c{}_d = i \left(\delta_a^c \, \eta_{bd} - \delta_b^c \, \eta_{ad} \right),
\]
and consequently
\be
\D_\mu \phi^a = \partial_\mu \phi^a + A^{a}{}_{b \mu} \, \phi^b.
\ee

Denoting a tetrad field by $h^a{}_\mu$, the spacetime metric $g_{\mu \nu}$ and the tangent-space Minko\-ws\-ki metric $\eta_{ab}$ are related through
\be
g_{\mu \nu} = \eta_{ab} \, h^a{}_\mu h^b{}_\nu.
\ee
Analogously, the linear connection $\Gamma^{\rho}{}_{\nu \mu}$ corresponding to $A^{a}{}_{b \mu}$ is
\be
\Gamma^{\rho}{}_{\nu \mu} = h_{a}{}^{\rho} \partial_{\mu} h^{a}{}_{\nu} +
h_{a}{}^{\rho} A^{a}{}_{b \mu} h^{b}{}_{\nu} \equiv h_{a}{}^{\rho} \D_{\mu} h^{a}{}_{\nu}.
\label{geco}
\ee
The inverse relation is
\be
A^{a}{}_{b \mu} =
h^{a}{}_{\nu} \partial_{\mu}  h_{b}{}^{\nu} +
h^{a}{}_{\nu} \Gamma^{\nu}{}_{\rho \mu} h_{b}{}^{\rho} \equiv h^{a}{}_{\nu} \nabla_{\mu} h_{b}{}^{\nu},
\label{gsc}
\ee
with $\nabla_{\mu}$ the covariant derivative defined by the connection $\Gamma^{\nu}{}_{\rho \mu}$. Equations (\ref{geco}) and (\ref{gsc}) are different ways of expressing the property that the total covariant derivative---that is, with connection terms for both indices---of the tetrad vanishes
identically:
\be
\partial_{\mu} h^{a}{}_{\nu} - \Gamma^{\rho}{}_{\nu \mu} h^{a}{}_{\rho} +
A^{a}{}_{b \mu} h^{b}{}_{\nu} = 0.
\label{todete}
\ee

The curvature and torsion of $A^{a}{}_{b \mu}$ are defined re\-spec\-tive\-ly by
\be
R^a{}_{b \nu \mu} = \partial_{\nu} A^{a}{}_{b \mu} -
\partial_{\mu} A^{a}{}_{b \nu} + A^a{}_{e \nu} A^e{}_{b \mu}
- A^a{}_{e \mu} A^e{}_{b \nu}
\ee
and
\be
T^a{}_{\nu \mu} = \partial_{\nu} h^{a}{}_{\mu} -
\partial_{\mu} h^{a}{}_{\nu} + A^a{}_{e \nu} h^e{}_{\mu}
- A^a{}_{e \mu} h^e{}_{\nu}.
\label{tordef}
\ee
Using relations (\ref{geco}) and (\ref{gsc}), they can be expressed in a purely spacetime form as
\be
\label{sixbm}
R^\rho{}_{\lambda\nu\mu} = \partial_\nu \Gamma^\rho{}_{\lambda \mu} -
\partial_\mu \Gamma^\rho{}_{\lambda \nu} +
\Gamma^\rho{}_{\eta \nu} \Gamma^\eta{}_{\lambda \mu} -
\Gamma^\rho{}_{\eta \mu} \Gamma^\eta{}_{\lambda \nu}
\ee
and
\be
T^\rho{}_{\nu \mu} =
\Gamma^\rho{}_{\mu\nu}-\Gamma^\rho{}_{\nu\mu}.
\label{sixam}
\ee
A Lorentz connection $\Gamma^\rho{}_{\mu\nu}$ can always be decomposed in the form\footnote{All quantities related to general relativity will be denoted with an over ``$\circ$''.}
\be
\Gamma^\rho{}_{\mu\nu} = {\stackrel{\circ}{\Gamma}}{}^{\rho}{}_{\mu \nu} +
K^\rho{}_{\mu\nu},
\label{prela0}
\ee
where
\be
{\stackrel{\circ}{\Gamma}}{}^{\sigma}{}_{\mu \nu} = {\textstyle
\frac{1}{2}} g^{\sigma \rho} \left( \partial_{\mu} g_{\rho \nu} +
\partial_{\nu} g_{\rho \mu} - \partial_{\rho} g_{\mu \nu} \right)
\label{lci}
\ee
is the Levi-Civita connection and
\be
K^\rho{}_{\mu\nu} = {\textstyle
\frac{1}{2}} \left(T_\nu{}^\rho{}_\mu + T_\mu{}^\rho{}_\nu -
T^\rho{}_{\mu\nu}\right) 
\label{contor}
\ee
is the contortion tensor. Using relation (\ref{geco}), the decomposition (\ref{prela0}) can be rewritten as
\be
A^a{}_{b\nu} = \Abol^a{}_{b\nu} + K^a{}_{b\nu},
\label{rela00}
\ee
where $\Abol^a{}_{b \nu}$ is the spin connection of general relativity. It is important to remark that, under a local Lorentz transformation
\be
h^a = \Lambda^a{}_b(x) \, h'^b,
\ee
the connection, $A^{a}{}_{b \mu}$ changes according to
\be
A^{a}{}_{b \mu} = \Lambda^{a}{}_{c}(x) \, A'^{c}{}_{d \mu} \,
\Lambda_{b}{}^{d}(x) + \Lambda^{a}{}_{c}(x) \, \partial_{\mu} \Lambda_{b}{}^{c}(x).
\label{ltsc}
\ee
If a connection vanishes in a given frame, therefore, it will be non-vanishing in a Lorentz-rotated frame. Since the last index of the connection is a tensorial index, one can write
\be
A^{a}{}_{b c} = A^{a}{}_{b \mu} \, h_c{}^\mu.
\label{Aabc}
\ee 

In the tetrad formalism, the quantities
\be
h^a = h^a{}_\mu \, dx^\mu \quad \mbox{and} \quad
h_a = h_a{}^\mu \, \partial_\mu
\ee
represent Lorentz frames. These frames are classified according to the value of the coefficient of anholonomy $f^{c}{}_{a b}$, which is de\-fined by the commutation relation \cite{livro}
\begin{equation}
[h_{a}, h_{b}] = f^{c}{}_{a b}\ h_{c}.
\label{eq:comtable}
\end{equation}
As a simple calculation shows, it is given by
\be
f^a{}_{cd} = h_c{}^\mu \, h_d{}^\nu (\partial_\nu
h^a{}_\mu - \partial_\mu h^a{}_\nu).
\ee
For example, the class of inertial frames is defined by all frames for which $f^{c}{}_{a b} = 0$. They are called, for this reason, holonomic frames. Starting from an inertial frame, different classes of frames are obtained by performing {\em local} (point dependent) Lorentz transformations. Inside each class, different frames are related through {\em global} (point independent) Lorentz transformations.

In special relativity, the anholonomy of the frames is entirely related to the inertial forces present in those frames. The preferred class of inertial frames is characterized by the absence of inertial forces, and consequently by holonomic frames. Similarly, in the presence of gravitation there is also a preferred class of frames: the class whose anholonomy is related to gravitation only, not with inertial effects. This class of frames, therefore, reduces to the inertial class when gravitation is switched off.

\section{Gravitational field equations and conservation laws}
\label{NecSuf}

In the first order formalism, where the lagrangian is assumed to depend on the tetrad and on its first derivative only, the gravitational field equation can be obtained from the usual Euler-Lagrange equation
\be
\frac{\partial {\mathcal L}}{\partial h^a{}_{\rho}} -
\partial_\sigma \frac{\partial {\mathcal L}}{\partial (\partial_\sigma h^a{}_{\rho})} = 0.
\ee
Denoting $h = \sqrt{-g}$, {with} $h = \det(h^a{}_\mu)$ {and} $g = \det(g_{\mu \nu})$, and introducing the constant $k = 8 \pi G/c^4$, the sourceless gravitational field equation can be written in the form
\be
\partial_{\sigma} (h S_a{}^{\rho \sigma}) - k \,
 h {\jmath}_a{}^{\rho} = 0,
\label{gfe}
\ee
where
\be
S_a{}^{\rho \sigma} = - \, S_a{}^{\sigma \rho} \equiv - \,
\frac{k}{h} \, \frac{\partial {\mathcal L}}{\partial (\partial_\sigma h^a{}_{\rho})}
\ee
is the so-called {\em superpotential}, and
\be 
\jmath_a{}^{\rho} \equiv - \, \frac{1}{h} \frac{\partial {\mathcal L}}{\partial h^a{}_{\rho}}  
\label{gcu}
\ee     
stands for the gravitational energy-momentum current. Equation (\ref{gfe}) is known as the {\it potential form} of the gravitational field equation \cite{Moller}. It is, in structure, similar to the Yang-Mills equation. Its main virtue is to explicitly exhibit the complex defining the energy-momentum current of the gravitational field. Due to the anti-symmetry of $S_a{}^{\rho \sigma}$ in the last two indices, the field equation implies the conservation of the gravitational energy-momentum current:
\be
\partial_\rho (h \jmath_a{}^\rho) = 0.
\label{conser1}
\ee

Any conservation law of this form, namely, written as a four-dimensional {\it ordinary} divergence, is a {\em true conservation law} in the sense that it yields a time-conserved {\em charge}---provided, of course, $\jmath_a{}^i$ ($i=1,2,3$) vanishes at infinity. On the other hand, in order to be physically meaningful, the equation expressing a conservation law must be covariant under both general coordinate and local Lorentz transformations. This simple property has an important consequence: {\em no tensorial quantity can be truly conserved}. In fact, since the derivative is not covariant, in order to yield a covariant conservation law, the conserved current cannot be covariant either. This means that the energy-momentum current $\jmath_a{}^\rho$ appearing in the gravitational field equation (\ref{gfe}) cannot be a tensor. Conversely, since the symmetric energy-momentum tensor of any source field---that is, the energy-momentum tensor appearing in the right-hand side of the gravitational field equation---is a true tensor, it is conserved in the covariant sense, otherwise the conservation law itself would not be covariant.

If a {\em tensorial} expression $t_a{}^{\rho}$ for the energy-momentum density of gravi\-tation exists, therefore, it has to be conserved in the covariant sense. On the other hand, due to the local Lorentz invariance of gravitation, inertial effects related to the non-inertiality of the frame will in general be present. These effects, as is well known, are non-covariant by its very nature. As a consequence, the energy-momentum density associated to the inertial effects, which we denote by $\imath_a{}^\rho$, cannot be a true tensor. This means that the sum of the inertial and the purely gravitational energy-momentum densities, $\imath_a{}^\rho + t_a{}^{\rho}$, will be a pseudotensor. This is actually a matter of consistency: since it represents the total energy-momentum density (in the absence of source fields), it has to be truly conserved,
\be
\partial_\rho [h (\imath_a{}^\rho + t_a{}^{\rho})] = 0,
\ee
and consequently cannot be a true tensor. Let us see next what happens in the specific case of general relativity.

\section{General relativity}

As is well known, it is not possible to construct an {\em invariant} la\-gran\-gi\-an for general relativity in terms of the metric and its first derivatives. There is, however, the second-derivative Einstein-Hilbert invariant lagrangian, in which the second-derivative terms reduce to a total divergence. This lagrangian can be written in the form 
\be
\Lbol = -\frac{h}{2 k} \, \Rbol \equiv \Lbol_{_1} + \partial_\mu (h \, w^\mu),
\ee
where $\Lbol_{_1}$ is a first-order lagrangian and $w^\mu$ is a contravariant four-vector. There are actually infinitely many different first-order la\-gran\-gi\-ans, each one connected to a different surface term:
\be
\Lbol = \Lbol_{_1} + \partial_\mu (h \, w^\mu) = \Lbol'_{_1} + \partial_\mu (h \, w'^\mu) = \Lbol''_{_1} +
\partial_\mu (h \, w''^\mu) = \dots \; .
\ee
Since the divergence term does not contribute to the field equation, any one of the first order lagrangians yields a field equation of the form
\begin{equation}
\partial_\sigma (h \Sbol_a{}^{\rho \sigma}) -
k \, h \jbol_a{}^\rho = 0,
\label{eqs1bis}
\end{equation}
which is the so-called potential form of Einstein's field equation \cite{Moller}. In this equation, $\Sbol_a{}^{\rho \sigma} = - \Sbol_a{}^{\sigma \rho}$ is the superpotential and $\jbol_a{}^{\rho}$ is the gravitational current, which is con\-served in the ordinary sense:
\be
\partial_\rho (h \jbol_a{}^\rho) = 0.
\ee
We see in this way that different first-order lagrangians give rise to different expressions for the gravitational energy-momentum pseudotensor.

Now, general relativity is strongly grounded on the equivalence principle, which says that it is always possible to find a local frame in which $\Abol^a{}_{b \mu} = 0$ along an observer world-line. In other words, at each point of the world-line, inertia compensates gravitation yielding a vanishing Lorentz connection. This means essentially that inertial and gravitational effects are both embodied in the spin connection $\Abol^a{}_{b \mu}$, and cannot be separated because of the equivalence principle. As a consequence of this inseparability, the gravitational energy-momentum current in general relativity will always include, in addition to the purely gravitational density, also the energy-momentum density of inertia. Since the latter is a pseudotensor, the whole current will also be a pseudotensor. In general relativity, therefore, it is not possible to define a tensorial expression for the gravitational energy-momentum density. This is in a\-gree\-ment with the strong equivalence principle which precludes the existence of such definition \cite{mtw}.

\section{Separating gravitation from inertia}
\label{sIfG}

In the class of frames $h'_b$ which reduces to inertial class when gravitation is switched off, the spin connection of teleparallel gravity\footnote{All quantities related to teleparallel gravity will be indicated by an over ``$\bullet$''.} vanishes eve\-ry\-where 
\be
\Aw'^{a}{}_{b \mu} = 0.
\label{wsc}
\ee
In these frames, the tetrad has the form \cite{review}
\be
h'^a{}_\mu = \partial_\mu x'^a + B'^a{}_\mu,
\ee
where $B'^a{}_\mu$ is the translational gauge potential. Observe that, if gravitation is switched off,  the translational gauge potential is of the form $B'^a{}_\mu = \partial_\mu \alpha^a(x)$, and the tetrad becomes trivial, or holonomic. In a Lorentz rotated frame, the tetrad assumes the form
\be
h^a{}_\mu = \Lambda^a{}_b(x) \, h'^b{}_\mu \equiv \partial_\mu x^a + \Aw^a{}_{b \mu} x^b + B^a{}_\mu,
\ee
where $B^a{}_\mu = \Lambda^a{}_b(x) \, B'^b{}_\mu$, and
\be
\Aw^{a}{}_{b \mu} = \Lambda^{a}{}_{e}(x) \, \partial_\mu \Lambda_{b}{}^{e}(x)
\label{ltwsc}
\ee
is the transformed connection. We see clearly from this expression that it represents purely inertial effects, not gravitation. Its curvature, consequently, vanishes identically: $\Rw^a{}_{b \mu \nu} = 0$. For a non-trivial ($B^a{}_\mu \neq 0$) tetrad, however, it has non-vanishing torsion: $\Tw^a{}_{\mu \nu} \neq 0$. In terms of the coefficient of anholonomy, it is given by
\be
\Tw^a{}_{bc} = - f^a{}_{bc} - (\Aw^a{}_{bc} - \Aw^a{}_{cb}).
\label{TfAA}
\ee

Let us consider now the equation of motion of a free particle in Minkowski spacetime. Seen from a holonomic frame $h'_b$, it is given by
\be
\frac{d u'^a}{ds} = 0.
\label{EqMot0}
\ee
Seen from an anholonomic frame $h_b$, which satisfies the commutation relation (\ref{eq:comtable}), it assumes the form \cite{MosPe}
\be
\frac{d u^a}{ds} + \onehalf \left(f_{b}{}^a{}_{c} + f_{c}{}^a{}_{b}
- f^{a}{}_{b c}\right) u^b \, u^c = 0,
\label{EqMot1}
\ee
where the coefficient
$
\onehalf \left(f_{b}{}^a{}_{c} + f_{c}{}^a{}_{b}
- f^{a}{}_{b c}\right)
$
represents a purely inertial connection, usually called Ricci coefficient of rotation.

Now, in order to obtain the corresponding equation of motion in the presence of gravitation, it is necessary to use the strong equivalence principle. Namely, if inertial and gravitational effects are locally {\it equivalent}, we can then assume that the above inertial connection represents the spin connection of general relativity \cite{weinberg}, 
\begin{equation}%
\onehalf \left(f_{b}{}^a{}_{c} + f_{c}{}^a{}_{b}
- f^{a}{}_{b c}\right) = \Abol^{a}{}_{b c}. 
\label{tobetaken3}
\end{equation}%
In this case, the equation of motion (\ref{EqMot1}) turns out to be the geodesic equation
\be
\frac{d u^a}{ds} + \Abol^{a}{}_{b c} \, u^b \, u^c = 0,
\label{EqMot2}
\ee
that is, the equation of motion of a spinless particle in the presence of gravitation. As already discussed, both inertia and gravitation are included in the connection $\Abol^{a}{}_{b c}$ and cannot be separated \cite{inertial}. This is a fundamental characteristic of the geometric approach of general relativity.

On the other hand, in the specific case of teleparallel gravity, the equation of motion of a (spinless) particle in the presence of gravitation is obtained from the free equation of motion (\ref{EqMot1}) by identifying
\begin{equation}%
\onehalf \left(f_{b}{}^a{}_{c} + f_{c}{}^a{}_{b}
- f^{a}{}_{b c}\right) = \Aw^{a}{}_{b c} - \Kw^{a}{}_{b c}.
\label{TeleIdent}
\end{equation}%
The equation of motion in this case assumes the form of a force equation,
\be
\frac{d u^a}{ds} + \Aw^{a}{}_{b c} \, u^b \, u^c =
\Kw^{a}{}_{b c} \, u^b \, u^c,
\label{EqMot4}
\ee
with contortion playing the role of gravitational force. This is the teleparallel version of the particle equation of motion. Of course, due to identity \cite{equiva}
\be
\Abol^{a}{}_{b c} = \Aw^{a}{}_{b c} - \Kw^{a}{}_{b c},
\label{splitting}
\ee
it is equivalent to the geodesic equation (\ref{EqMot2}).

In its standard formulation, the strong equivalence principle says that it is always possible to find a frame in which inertia compensates gravitation {\it locally}, that is, in a point or along a world-line. In that local frame, the spin connection of general relativity vanishes:
\be
\Abol^{a}{}_{b c} = 0.
\ee
On account of the identity (\ref{splitting}), the teleparallel version of this expression is
\be
\Aw^{a}{}_{b c} = \Kw^{a}{}_{b c},
\label{A=K}
\ee
which shows explicitly that, in such frame, inertia (left-hand side) compensates gravitation (right-hand side). We see in this way that the splitting of the spin connection according to equation (\ref{splitting}) corresponds actually to a separation between inertia and gravitation \cite{Einstein05}. As a consequence, the right-hand side of the equation of motion (\ref{EqMot4}) represents the purely gravitational force, and transforms co\-var\-iantly under local Lorentz transformations. The inertial effects coming from the frame non-inertiality are represented by the connection term of the left-hand side, which is non-covariant by its very nature. In teleparallel gravity, therefore, whereas the gravitational effects are described by a covariant force \cite{paper1}, the non-inertial effects of the frame remain {\it geometrized} in the sense of general relativity, and are represented by an inertial-related connection.

It is important to remark that, because inertial and gravitational effects work now separately, teleparallel gravity does not require the equivalence between inertia and gravi\-ta\-tion to describe the gravitational interaction. In other words, it does not require the strong equivalence principle \cite{DoWiEP}. To see that this is in fact the case, we recall that it is also possible to choose a {\it global} frame $h'_b$ in which the inertial effects vanish. In this frame, $\Aw'^{a}{}_{b c} = 0$ and the equation of motion (\ref{EqMot4}) becomes purely gravitational:
\be
\frac{d u'^a}{ds} = \Kw'^a{}_{b c} \, u'^b \, u'^c.
\label{eqmot3}
\ee
Since no inertial effects are present, it becomes clear that teleparallel gravity does not need to use the equivalence between inertia and gravitation. In this form, the gravitational force becomes quite similar to the Lorentz force of electrodynamics. {There is a crucial difference, though: due to the strict attractive character of gravitation, the gravitational force is quadratic in the four-velocity, in contrast to the Lorentz force of electrodynamics, which is linear in the four-velocity.}

\section{Purely gravitational energy-momentum density}

The gravitational lagrangian of teleparallel gravity is \cite{maluf}
\be
\Lw = \frac{h}{4 k} \; \sw_{a}{}^{\mu \nu} \; \Tw^a{}_{\mu \nu},
\label{tela}
\ee
where
\be
\sw_a{}^{\rho \sigma} \equiv - \,
\frac{k}{h} \, \frac{\partial {\Lw}}{\partial (\partial_\sigma h^a{}_{\rho})} = \Kw^{\rho \sigma}{}_{a} - h_{a}{}^{\sigma} \;  \Tw^{\theta \rho}{}_{\theta} + h_{a}{}^{\rho} \; \Tw^{\theta \sigma}{}_{\theta}
\label{supote}
\ee
is the superpotential, with $\Kw^{\rho \sigma}{}_{a}$ the teleparallel contortion. Seen from a general frame $h_a$, the corresponding sourceless gravitational field equation, which in tele\-par\-al\-lel gravity is naturally written in the potential form, is given by 
\be
\partial_\sigma (h \sw_a{}^{\rho \sigma}) -
k \, h \, \jw_{a}{}^{\rho} = 0,
\label{tfe0}
\ee
where
\be
\jw_{a}{}^{\rho} \equiv - \, \frac{1}{h} \frac{\partial {\Lw}}{\partial h^a{}_{\rho}} =
\frac{1}{k} \, h_a{}^{\lambda} \, \sw_c{}^{\nu \rho} \,
\Tw^c{}_{\nu \lambda} - \frac{h_a{}^{\rho}}{h} \, \Lw +
\frac{1}{k} \, \Aw^c{}_{a \sigma} \sw_c{}^{\rho \sigma}
\label{ptem10bis}
\ee
is the teleparallel energy-momentum current. Due to the anti-symmetry of the su\-per\-potential in the last two indices, it is conserved in the ordinary sense:
\be
\partial_\rho (h \jw_{a}{}^{\rho}) = 0.
\label{iplustCon}
\ee

Now, since the teleparallel spin connection $\Aw^c{}_{a \sigma}$ represents inertial effects, the last term of the current (\ref{ptem10bis}) can be interpreted as the energy-momentum density of the inertial forces: 
\be
\iw_{a}{}^{\rho} = \frac{1}{k} \, \Aw^c{}_{a \sigma} \sw_c{}^{\rho \sigma}.
\label{InerEM}
\ee
Of course, by its very nature, it is non-covariant. The total (grav\-i\-tation plus inertia) en\-er\-gy-momentum density is consequently
\be
\jw_{a}{}^{\rho} = \tw_{a}{}^{\rho} + \iw_{a}{}^{\rho},
\ee
where
\be
\tw_{a}{}^{\rho} =
\frac{1}{k} \, h_a{}^{\lambda} \, \sw_c{}^{\nu \rho} \,
\Tw^c{}_{\nu \lambda} - \frac{h_a{}^{\rho}}{h} \, \Lw
\label{graem}
\ee
is a {\em tensorial current which represents the energy-momentum density of gravity}. If considered separately from inertia, therefore, the gravitational energy-momentum density is found to be a true tensor. As far as teleparallel gravity does not require the equivalence principle to describe the gravitational interaction (see section~\ref{sIfG}), the ex\-ist\-ence of a tensorial expression for the gravitational energy-momentum density is not in conflict with the equivalence principle \cite{gemt}.

Let us now obtain the conservation law of the energy-momentum tensor $\tw_{a}{}^{\rho}$. To begin with, we observe that the potential term of field equation (\ref{tfe0}) and the inertial current $\iw_{a}{}^{\rho}$ make up a Fock-Ivanenko covariant derivative of the superpotential:
\be
\partial_\sigma (h \sw_a{}^{\rho \sigma}) -
\Aw ^c{}_{a \sigma} (h \, \sw_{c}{}^{\rho}) \equiv \Dw_\sigma (h \sw_a{}^{\rho \sigma}).
\label{DotFI}
\ee
This allows us to rewrite that field equation in the form
\be
\Dw_\sigma (h \sw_a{}^{\rho \sigma}) -
k \, h \, \tw_{a}{}^{\rho} = 0,
\label{fe11}
\ee
Since the teleparallel spin con\-nec\-tion (\ref{ltwsc}) has vanishing curvature, the corresponding Fock-Ivanenko derivative is com\-mutative:
\be
[\Dw_\rho , \Dw_\sigma] = 0.
\ee
Taking into account the anti-symmetry of the su\-per\-potential in the last two indices, it follows from the field equation (\ref{fe11}) that the tensorial current (\ref{graem}) is conserved in the  covariant sense:
\be
\Dw_\rho (h \tw_{a}{}^{\rho}) = 0.
\label{GravEMcon}
\ee
Of course, as it does not represent the total energy-momentum density---in the sense that the inertial part is not included---it does not need to be truly conserved.

It is important to notice finally that, in a locally inertial frame, $f^a{}_{bc} = 0$, and we see from from Eq.~(\ref{TfAA}) that
\be
\Tw^a{}_{bc} = - (\Aw^a{}_{bc} - \Aw^a{}_{cb}),
\label{TfAAmodi}
\ee
which is equivalent to the local identity (\ref{A=K}). Similarly, the superpotential can be show to satisfy a similar local relation between gravitation and inertia,
\be
\sw_a{}^{bc} = \siw_a{}^{bc},
\label{Id}
\ee
where
\be
\siw_a{}^{bc} = \Aw^{bc}{}_a + \delta_a^c \, \Aw^{eb}{}_e - \delta_a^b \, \Aw^{ec}{}_e
\ee
is a purely inertial superpotential. In absence of gravitation, therefore, we can use identity (\ref{Id}) to rewrite the inertial energy-momentum pseudotensor in the form
\be
\iw_{a}{}^{\rho} = \frac{1}{k} \, \Aw^c{}_{a \mu} \, \siw_c{}^{\rho \mu},
\ee
where we have used the notation $\siw_c{}^{\rho \mu} = \siw_c{}^{de} \, h_d{}^\rho h_e{}^\mu$. This is the energy-momentum pseudotensor of inertia.

\section{Concluding remarks}

In general relativity, both gravitational and inertial effects are mixed in the connection of the theory, and cannot be separated. Even though some quantities, like curvature, are not affected by inertia, some others turn out to depend on it. For example, the energy-momentum density of gravitation will necessarily include both the energy-momentum density of gravity and the energy-momentum density of inertia. Since the in\-er\-tial effects are non-tensorial---they depend on the frame---the complex defining the energy-momentum density of the gravitational field in general relativity always shows up as a non-tensorial object.

On the other hand, although equivalent to general relativity, teleparallel gravity is able to separate gravitation from inertia \cite{Einstein05}. As a consequence, it becomes possible to write down an energy-momentum density for gravitation only, excluding the contribution from inertia. In the preferred class of frames $h'_a$, characterized by the absence of inertial effects, the teleparallel spin connection vanishes, $\Aw'^a{}_{b \mu} = 0$, and the gravitational field equation assumes the form
\be
\partial_\sigma (h \sw'_a{}^{\rho \sigma}) -
k \, h \, \tw'_{a}{}^{\rho} = 0,
\label{fe12}
\ee
with $\tw'_{a}{}^{\rho}$ the purely gravitational energy-momentum tensor. In this class of frames, the gravitational energy-momentum tensor is conserved in the ordinary sense:
\be
\partial_\rho (h \tw'_{a}{}^{\rho}) = 0.
\ee
The gravitational energy and momentum can then be obtained by integrating the appropriate components of $\tw'_{a}{}^{\rho}$ in the whole space-like volume $V$,
\be
P'_a = \int_V h \tw'_a,
\label{P1}
\ee
with $\tw'_a$ the gravitational energy-momentum tensor written in an appropriate basis.\footnote{For more details on this notation, see for example Ref.~\cite{basis}.} Alternatively, on account of the field equation (\ref{fe12}), they can be obtained by computing the surface integral of the superpotential at infinite,
\be
P'_a = \int_{\partial V} h S'_a,
\label{P2}
\ee
where $\partial V$ denotes the spatial boundary of $V$. Since in the frame $h'_a$ there is no inertial effects, this integral gives the correct result for the energy and momentum of gravitation.

Global Lorentz transformations $\Lambda_a{}^b$, as is well known, are related to frame changes inside a given class of frames. For example, all inertial frames of special relativity are related by global Lorentz transformations. This means that such transformations do not introduce inertial effects. Local Lorentz transformations $\Lambda_a{}^b(x)$, on the other hand, will change any frame of this class to a frame belonging to a different class, in which inertial effects are present. In this new class of frames, the teleparallel spin connection $\Aw^a{}_{b \mu}$ turns out to be that given by Eq.~(\ref{splitting}), and the field equation becomes
\be
\Dw_\sigma (h \sw_a{}^{\rho \sigma}) -
k \, h \, \tw_{a}{}^{\rho} = 0,
\label{fe11bis}
\ee
where\footnote{The superpotential $\sw_a{}^{\rho \sigma}$ is given by a similar transformation from $\sw'_b{}^{\rho \sigma}$.}
\be
\tw_{a}{}^{\rho} = \Lambda_a{}^b(x) \, \tw'_{b}{}^{\rho}.
\label{InerDep}
\ee
In this case, the gravitational energy-momentum tensor turns out to depend on the inertial effects present in the frame. Due to this dependence, its volume integration will not give the correct, or physically relevant result. As a matter of fact, this is true for the energy-momentum tensor of any field in special relativity: when transformed to a non-inertial frame, its volume integration does not give the correct energy-momentum density of the field because, even transforming covariantly, it includes also inertial effects. 

This problem can be circumvented by rewriting the field equation (\ref{fe11bis}) in the form
\be
\partial_\sigma (h \sw_a{}^{\rho \sigma}) -
k  h \, (\iw_{a}{}^{\rho} + \tw_{a}{}^{\rho}) = 0,
\label{fe13}
\ee
where $\iw_{a}{}^{\rho}$ is the inertial energy-momentum pseudotensor, given by Eq.~(\ref{InerEM}). Let us then  consider the integral
\be
P_a = \int_V h \, (\iw_{a}{}^{\rho} + \tw_{a}{}^{\rho}).
\label{j1}
\ee
On account of the field equation (\ref{fe13}), it can be rewritten as the surface integral of the superpotential at infinity:
\be
P_a = \int_{\partial V} h \sw_a.
\label{j2}
\ee
Now comes the crucial point. If $\Lambda_a{}^b(x)$ reduces to the identity transformation at infinite, the computation of the gravitational energy-momentum on any frame will give the same result as that obtained in the class of frames $h'_a$, where no inertial effects are present. On the other hand, if the local transformation $\Lambda_a{}^b(x)$ reduces to a global transformation $\Lambda_a{}^b$ at infinite---which means that no inertial effects are introduced on the boundary $\partial V$---the gravitational four-momentum is found to be covariantly transformed between these two classes of frames by a global transformation:
\[
P_a = {\Lambda}_a{}^b P'_b.
\]

Summing up, in teleparallel gravity it is possible to choose a preferred class of frames in which inertial effects are absent. Since these effects are responsible for spoiling the computation of the energy-momentum density, the use of the preferred class of frames yields the correct physical result. In any other frame related to the first by a local Lorentz transformation which does not introduce inertial effects at the space infinity,\footnote{Of course, a non-trivial local Lorentz transformation at spatial infinite would change the class of observers on the boundary $\partial V$, and would spoil the calculation of $P_a$.} the correct energy and momentum are obtained by integrating the pseudo-current
\be
\jw_{a}{}^{\rho} = \iw_{a}{}^{\rho} + \tw_{a}{}^{\rho}
\label{RegCur}
\ee
in the whole volume, or equivalently, by integrating the corresponding superpotential at infinity. In a sense, this is similar to what happens in general relativity. The difference is that, because teleparallel gravity allows a separation between gravitation and inertia, it is possible to understand now why the integration of the pseudo-current (\ref{RegCur}) gives the physical result in any frame: the inertial term $\iw_{a}{}^{\rho}$ compensates exactly the {local} dependence of $\tw_{a}{}^{\rho}$ on inertia, always yielding the physically relevant result. It can, consequently, be interpreted as a kind of ``regularized'' energy-momentum current.\footnote{Similar regularizing procedures have been discussed recently in Refs.~\cite{reg1,reg2}.}

\section*{Acknowledgments}
The authors would like to thank Y. Obukhov for useful discussions. They would like to thank also FAPESP, CAPES and CNPq for partial financial sup\-port.


\end{document}